\documentclass{sigchi}
\usepackage{graphicx}
\usepackage{url}
\newenvironment{myquote}{\list{}{\leftmargin=0.2in\rightmargin=0.2in}\item[]}{\endlist}

\newcommand\hcancel[2][black]{\setbox0=\hbox{$#2$}%
\rlap{\raisebox{.45\ht0}{\textcolor{#1}{\rule{\wd0}{1pt}}}}#2} 
\usepackage{xcolor,cancel}

\def\pprw{8.5in}
\def\pprh{11in}

\begin{document}

\title{A Crowdsourcing Approach To Collecting Tutorial Videos --\\Toward Personalized Learning-at-Scale}
\numberofauthors{3} 
%
\author{
\alignauthor{Jacob Whitehill\\
\affaddr{Worcester Polytechnic Institute}\\
\affaddr{Worcester, MA, USA}\\
\email{jrwhitehill@wpi.edu}}\\
\alignauthor{Margo Seltzer\\
\affaddr{Harvard University}\\
\affaddr{Cambridge, MA, USA}\\
\email{margo@eecs.harvard.edu}}\\
%
%
}
\date{2 March 2016}

\maketitle
\begin{abstract}
We investigated the feasibility of crowdsourcing full-fledged tutorial videos
from ordinary people on the Web on how to solve  math problems related to logarithms. 
This kind of approach (a form of \emph{learnersourcing}
\cite{kim2014understanding,kim2014crowdsourcing}) to efficiently collecting tutorial videos and
other learning resources could be useful for realizing \emph{personalized learning-at-scale},
whereby students receive specific learning resources -- drawn from a large and diverse set -- that are
tailored to their individual and time-varying needs.
Results of our study, in which we collected 399 videos from 66 unique ``teachers'' on Mechanical Turk,
suggest that (1) approximately 100 videos -- over
$80\%$ of which are mathematically fully correct -- can be crowdsourced per week for \$5/video;  (2)
the crowdsourced videos exhibit significant diversity in terms of language style,
presentation media, and pedagogical approach; (3)
the average learning gains (posttest minus pretest score) associated with watching the videos
was stat.~sig.~higher than for a control video ($0.105$ versus $0.045$); and
(4) the average learning gains ($0.1416$) from watching the best tested crowdsourced videos  was
comparable  to the learning gains ($0.1506$) from watching 
a popular Khan Academy video on logarithms.
\end{abstract}

\section{Introduction \& Related Work}

The goal of \emph{personalized learning}, in which students' learning experiences are tailored to their individual and time-varying needs, has been been pursued by psychologists,
computer scientists, and educational researchers for over five decades. Over the years, personalized learning systems have taken various forms:
computer-aided instruction systems
in the 1960-1970s; intelligent tutoring systems in the 1980-2000s \cite{vanlehn2005andes,anderson1985intelligent,woolf2009affect}; web-based e-learning platforms
in 2000-2010s \cite{brusilovsky2003adaptive,hwang2010heuristic,chen2008intelligent}; and adaptive learning engines --  as developed by companies such as Knewton, Pearson, and
McGraw-Hill -- from 2010-present. From an abstract perspective, the common goal of all these technologies is to provide each student at each moment in time with specific
\emph{learning resources} -- e.g., illuminating tutorials of key concepts, edifying practice problems, helpful explanations of how to solve these problems, etc. --
that can help students to learn more effectively than they could with a one-size-fits-all instructional approach.

A key challenge when developing personalized learning systems is {\bf how to efficiently collect a set 
of learning resources} that are used to personalize instruction. 
Without a sufficiently large and diverse set of resources from which to draw, personalized learning may not offer much advantage over
traditional, single-path instruction.
Intelligent tutoring systems in particular, for which the empirical benefits of personalized learning are arguably strongest \cite{pane2013effectiveness,koedinger2007exploring},
can be extremely laborious to create, and a significant part of the effort that must be invested is in the creation of good explanations and hints \cite{murray2003overview}.
Moreover, in order to be maximally effective, personalized learning systems should consider \emph{interactions} between
the learner and the resources with which they interact:
these interactions could be based on shared demographics of the learner and the teacher (e.g., role model
effects \cite{paredes2014teacher,lim2015impact,dee2005teacher}), language complexity of the resource that is tuned to the proficiency of the learner \cite{haag2013second},
affective sentiment (e.g., enthusiasm \cite{kunter2008students}, humor \cite{ziv1988teaching}) of the resource that matches
the emotional state of the learner, and more. Unfortunately, 
as the number of possible interactions between learners and resources increases, the problem of how to collect a large and diverse enough
pool of resources becomes increasingly severe.

One recently proposed and promising approach to collecting and curating large volumes of educational resources is to
\emph{crowdsource} data from learners themselves. This process, sometimes known as \emph{learnersourcing},
has been used, for example, to identify which parts of lecture videos are confusing \cite{kim2014understanding}, and to
describe the key instructional steps \cite{kim2014crowdsourcing} and subgoals \cite{kim2013learnersourcing} of ``how-to'' videos.
More recently, learnersourcing has been used not only to annotate existing educational content, but also to create novel content
itself. In particular, \cite{williams2016axis} explored a crowdsourcing-based strategy toward personalized learning
in which learners were asked to author paragraphs of text explaining how to solve statistics problems. The explanations generated
by learners were found to be comparable in both learning benefit and rated quality to explanations produced by expert
instructors.

In this paper, we too explore an approach to efficiently collecting a large and diverse set of learning resources that is based on crowdsourcing.
However, in contrast to  \cite{williams2016axis}, in which short text-based explanations were gathered from learners who were already
engaged in a learning task, our work is concerned with asking ordinary people from a crowdsourcing web site to take on the role of a teacher
(which has been dubbed ``teachersourcing'' \cite{HeffernanEtAl2016})
and to create \emph{novel, full-fledged, video-based explanations} 
that provide worked examples \cite{booth2013using} of how to solve
a variety of mathematics problems that could potentially help math students to learn. In contrast to static text,
multimedia videos such as whiteboard animations  can help to focus students' attention on the most salient
parts of an explanation -- e.g., by pointing to a specific mathematical expression with the mouse pointer while talking.
Moreover, some students may find video to be more engaging than text, 
and there is preliminary evidence from the education literature that multimedia presentations lead to greater knowledge retention
compared to static text-based presentations \cite{turkay2016effects}.
We note that the effort involved for the ``teachers'' in creating these videos
is considerable -- often an hour or more
of total time according to self-reports by the participants in our study. It is thus unclear how many people on crowdsourcing websites such as Mechanical Turk would even respond
to such a task, and even less clear how useful such crowdsourced explanations might be in terms of helping students to learn.

This paper describes what we believe to be the first investigation into crowdsourcing entire tutorial videos from ordinary people on the Web.
In particular, the rest of the paper investigates the following research questions:
\begin{enumerate}
\item How can we design a crowdsourcing task to convince
ordinary people to create, for a modest amount of compensation, a \emph{novel} tutorial video (not just a link to an existing video)
that might realistically be used to help students learn? What is the throughput (videos/week) that we can attain, and how
many of these videos are mathematically correct?
\item What kinds of qualitative diversity -- e.g., presentation style, pegagogical approach, language style -- do the crowdsourced videos exhibit?
\item How effective are these videos in helping students learn about the subject matter they are supposed to explain? How do they
compare with a video produced by Khan Academy?
\end{enumerate}

\section{Experiment I: Crowdsourcing Videos}
\label{sec:crowdsourcing}
In this study we focused on crowdsourcing tutorial videos that explain how to simplify mathematical expressions
and solve equations involving \emph{logarithms}. Logarithms are well-suited for this study because
many people know what they are; many other people -- even those who once learned them many years ago -- do not; and
people who are not familiar with logartithms can still learn something useful about them in a small (say, less than 10 minutes)
amount of time. In particular, we 
chose 18 math problems (see Figure \ref{fig:problems}) that were given as part of a pre-test from another research project
that was conducted by \cite{salamanca2012characterizing} on how math tutors interact with their students in traditional classroom
settings. As we will discuss in Experiment II, we will later also use the post-test from that same study.
\begin{figure}
\small
\noindent\fbox{%
    \parbox{3in}{
		\subsection*{Basic Logarithms}
		Simplify:\\
		\begin{tabular}{ll}
		$\log_3 1 =$      \hspace{2cm}   &  $\log_9 1 =$ \\
		$\log 100 =$         &  $\log_{\frac{1}{5}} 125 =$ \\
		$\log_{10} 1000 =$   &  $\log_{\frac{1}{x}} x^2 =$ \\
		$\log_3 81 =$   &  $\log_w \frac{1}{w} =$ \\
		$\log_2 8 =$         &  $\log_{\frac{1}{2}} \frac{1}{4} =$ \\
		\end{tabular}
		\subsection*{Logarithms and Variables}
		Simplify:\\
		\begin{tabular}{ll}
		$\log_a a^2 =$    \hspace{1.8cm}   & $\log_x x^4 =$ \\
		$\log_4 4^{2b} =$  & $\log_{x-1} (x-1)^y =$ \\
		\end{tabular}
		\subsection*{Equations with Logarithms}
		Solve:\\
		\begin{tabular}{ll}
		$\log_3 (x-1) = 4$    \hspace{1.1cm}          &   $x \log_4 16 = 3$ \\
		$z \log_{10} \sqrt{10} = 4$   &   $y \log_{10} 1000 = 3$ \\
		\end{tabular}
	}
}
\caption{The 18 math problems that the crowdsourced teachers were asked to solve and explain in novel video tutorials.}
\label{fig:problems}
\normalsize
\end{figure}

\subsection{Participants}
The ``teachers'' in our study were adult (18 years or older) workers on 
Amazon Mechanical Turk.
All participants were first required to give informed consent (the experiment was approved by our institution's IRB 
(IRB15-0867) and also sign a video recording release form so that their video explanations can be used in subsequent experiments on learning.
Participants who completed the experiment received a payment of \$5.

\subsection{Apparatus}
When interacting with the Mechanical Turk, workers complete one or more Human Intelligence Tasks (``HITs'').
A synopsis of the HIT we posted is shown in the Appendix in Figure \ref{fig:HIT}.  In place of the text ``PROBLEM'',
one of the problems from Figure \ref{fig:problems} -- such as ``Simplify $\log_2 8$'' -- is shown to the worker;
it is then her/his job to create a video explaining how to solve this problem.
For each problem of the 18 problems, we solicited workers on the Mechanical Turk to produce a video to explain
how to solve this problem to a student. Teachers were allowed to create one video for multiple problems if they desired,
but not multiple videos for the same problem.

\subsection{Procedure}
The experiment was conducted as follows:
\begin{enumerate}
\item We asked the participant to answer a brief survey about their age, gender, level of education, and interest in mathematics.
\item In order to give the participant an idea of what we were looking for, we asked her/him to watch several examples
of what a good video explanation might look like; the examples we chose were popular videos from Youtube about long division and quadratic equations.
\item For the benefit of participants who chose to record their own handwriting, we provided explicit guidelines on handwriting quality and
showed good and bad examples of each.
\item We presented one of the 18 problems mentioned above and asked them to create and upload a video explaining how to solve it.
\item The participant uploaded her/his video and completed the HIT.
\end{enumerate}

\subsection{Dependent variables}
We measured (1) the number of participants (``teachers'') who created a tutorial video, (2) the average number of 
tutorial videos created by each participant, and (3) the fraction of submitted videos that were mathematically correct.

\subsection{Results}
\begin{figure*}
    \parbox{\textwidth}{
\begin{center}
Video 1\\\vspace{.1cm}
\includegraphics[width=1.5in]{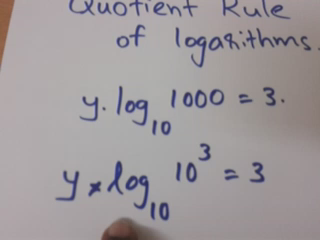}
\includegraphics[width=1.5in]{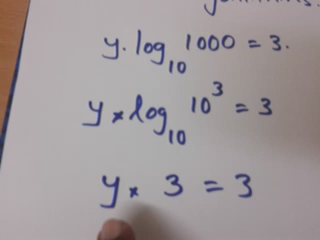}
\includegraphics[width=1.5in]{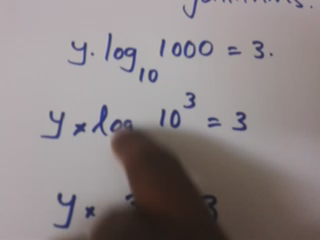}
\includegraphics[width=1.5in]{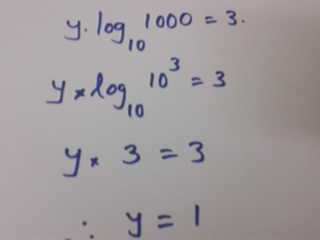}
\end{center}
    }

    \parbox{\textwidth}{
\begin{center}
Video 2\\\vspace{.1cm}
\includegraphics[width=1.5in]{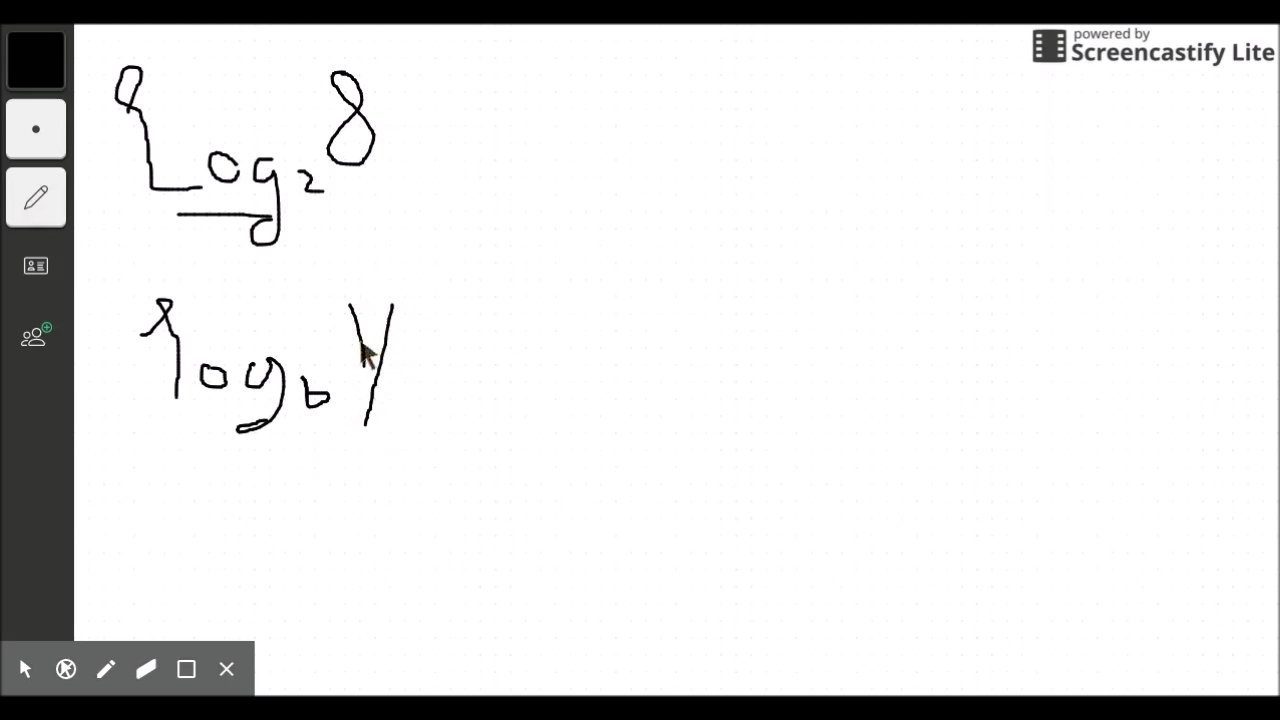}
\includegraphics[width=1.5in]{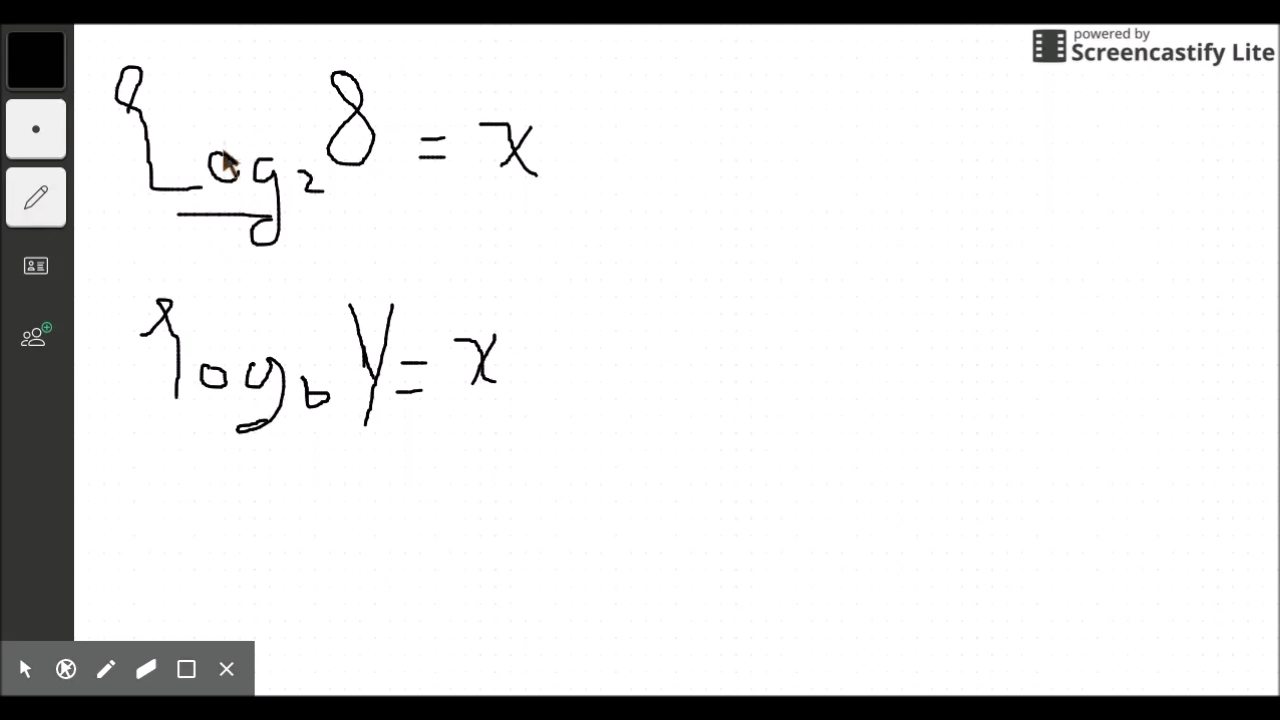}
\includegraphics[width=1.5in]{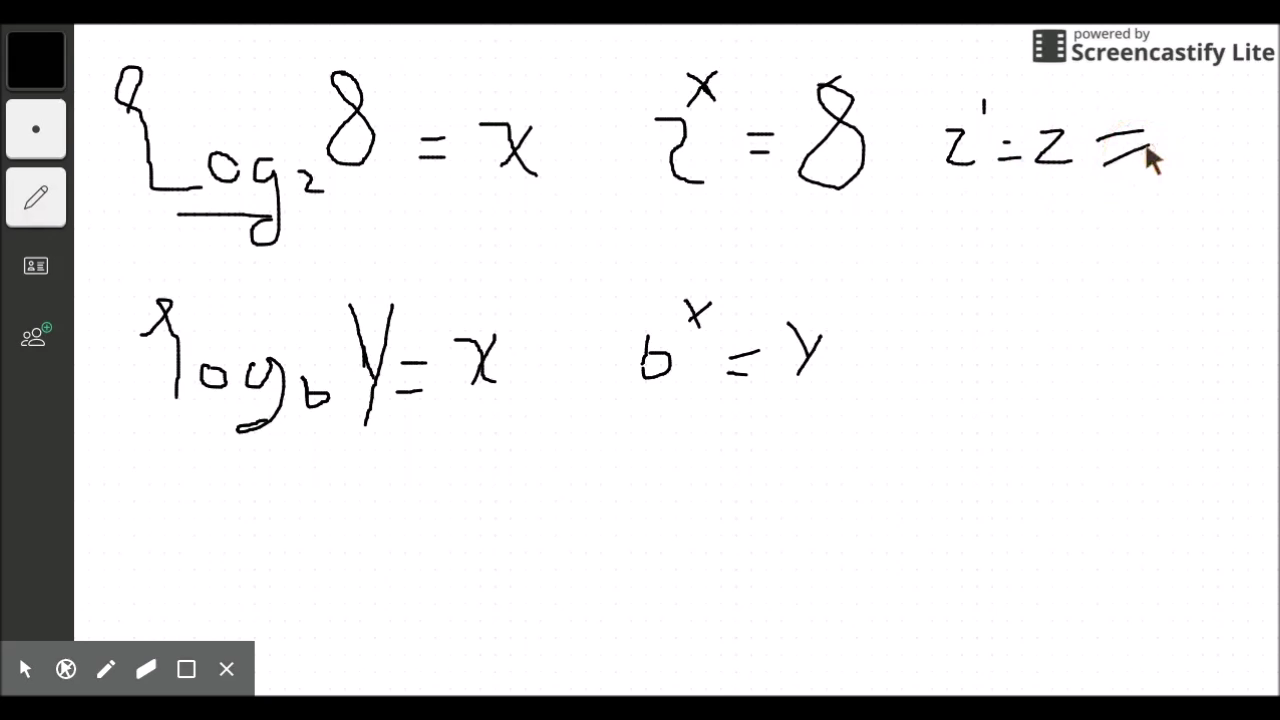}
\includegraphics[width=1.5in]{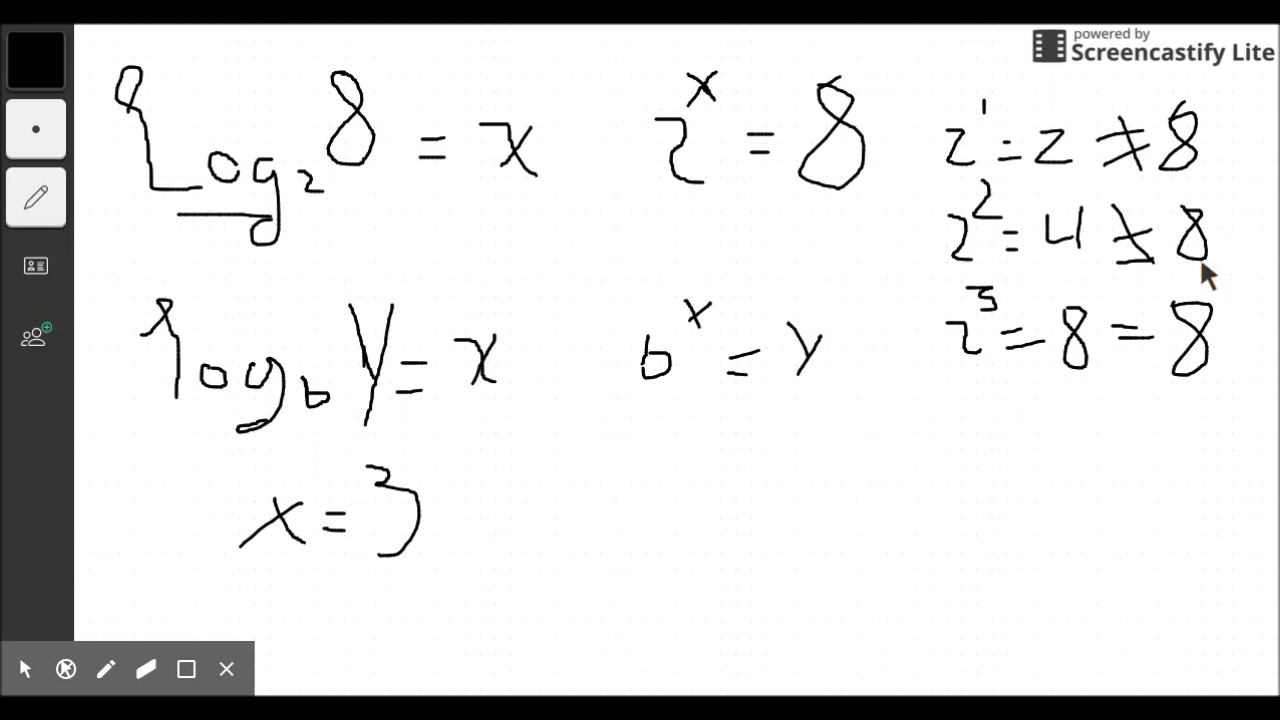}
\end{center}
    }

    \parbox{\textwidth}{
\begin{center}
Video 3\\\vspace{.1cm}
\includegraphics[width=1.5in]{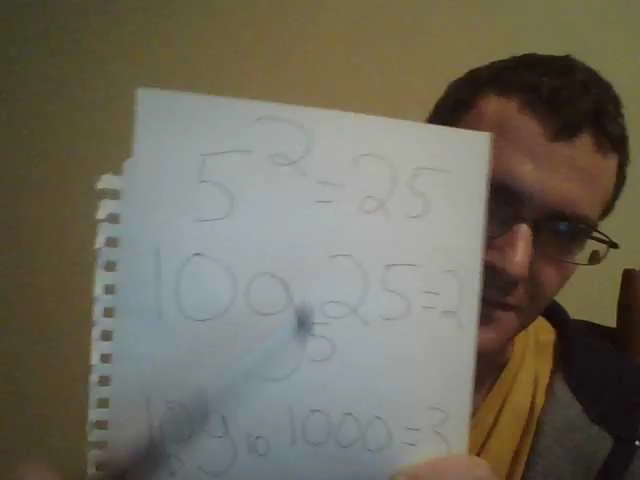}
\includegraphics[width=1.5in]{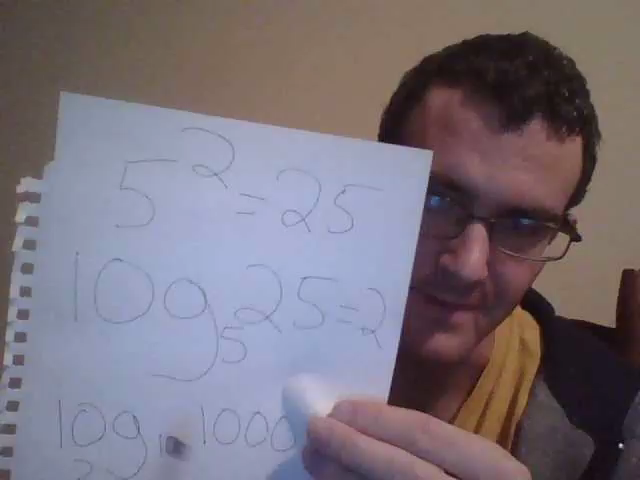}
\includegraphics[width=1.5in]{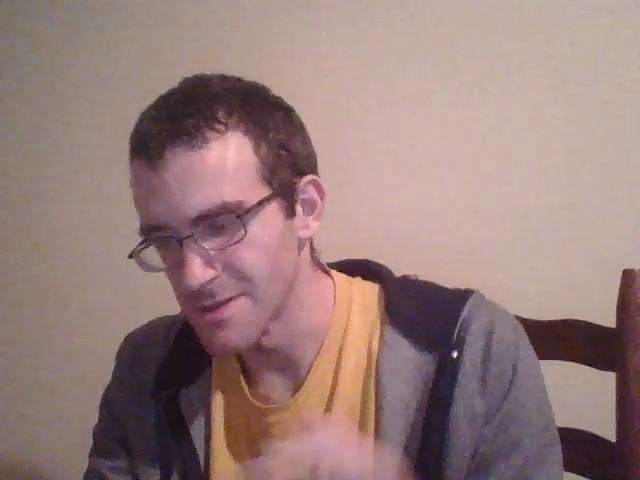}
\includegraphics[width=1.5in]{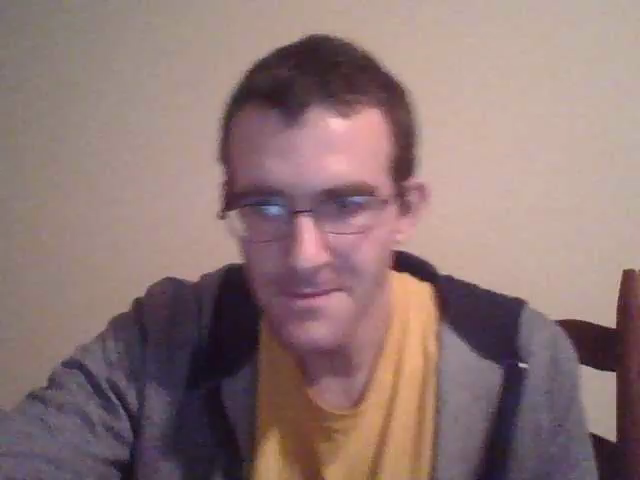}
\end{center}
    }

    \parbox{\textwidth}{
\begin{center}
Video 4\\\vspace{.1cm}
\includegraphics[width=1.5in]{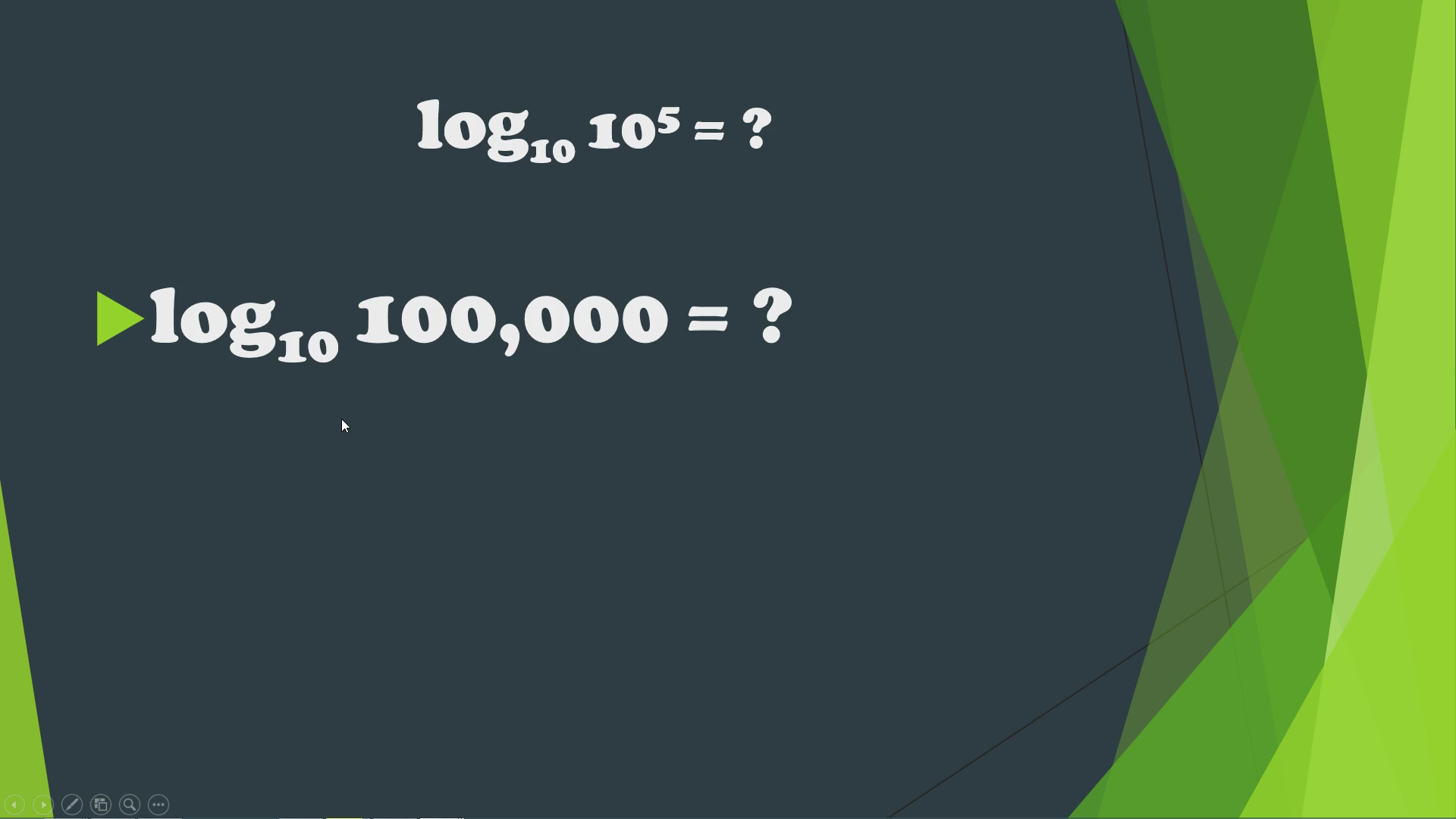}
\includegraphics[width=1.5in]{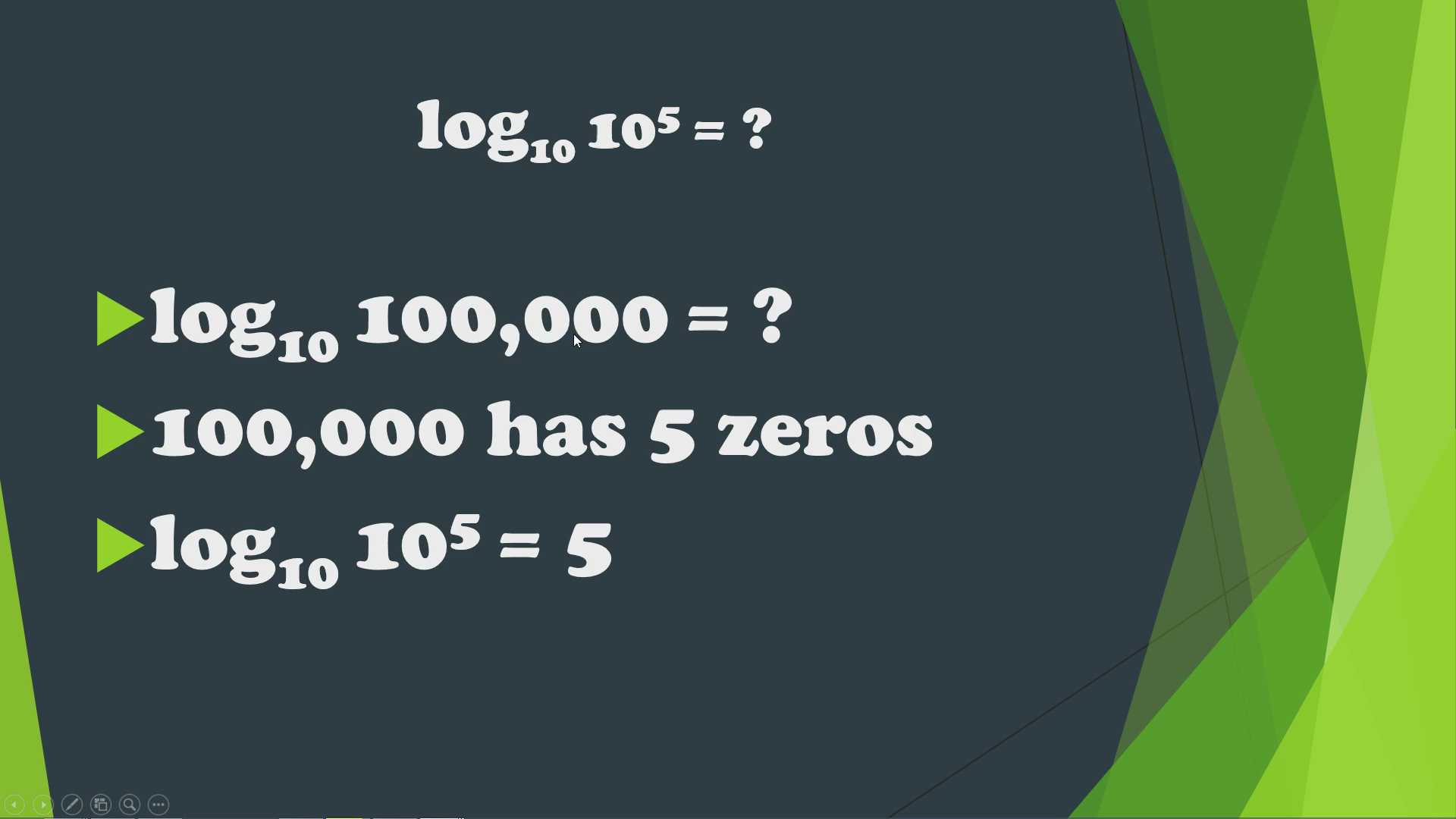}
\includegraphics[width=1.5in]{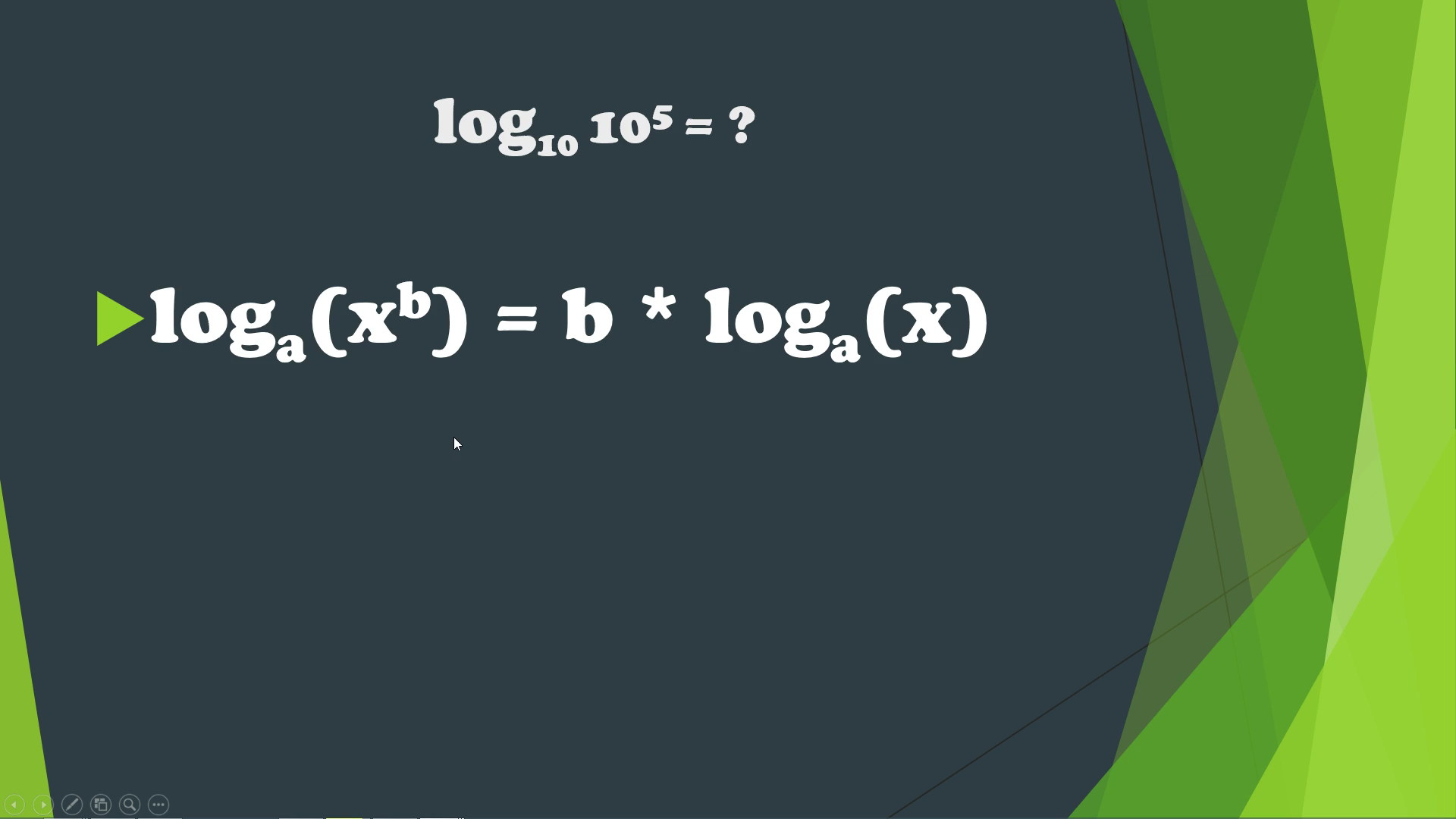}
\includegraphics[width=1.5in]{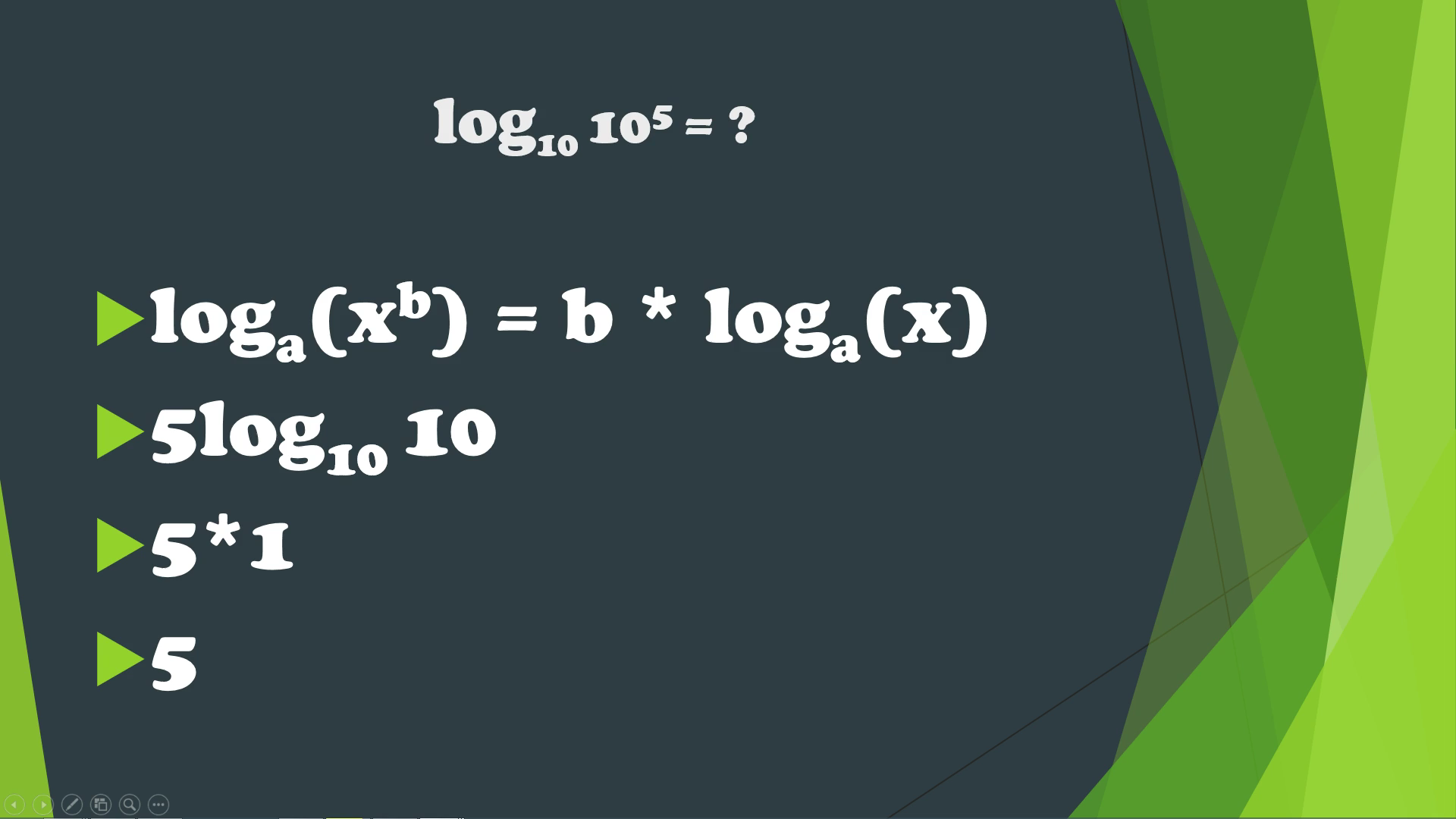}
\end{center}
    }

    \parbox{\textwidth}{
\begin{center}
Video 5\\\vspace{.1cm}
\includegraphics[width=1.65in]{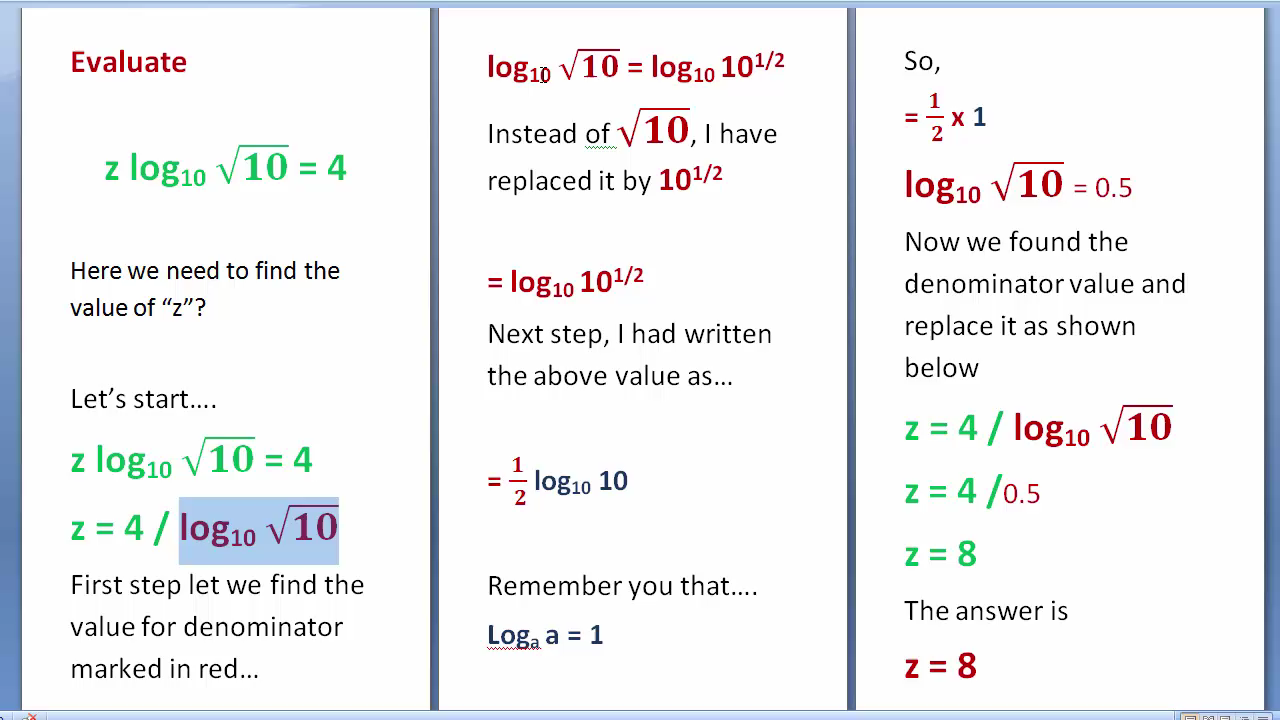}
\includegraphics[width=1.65in]{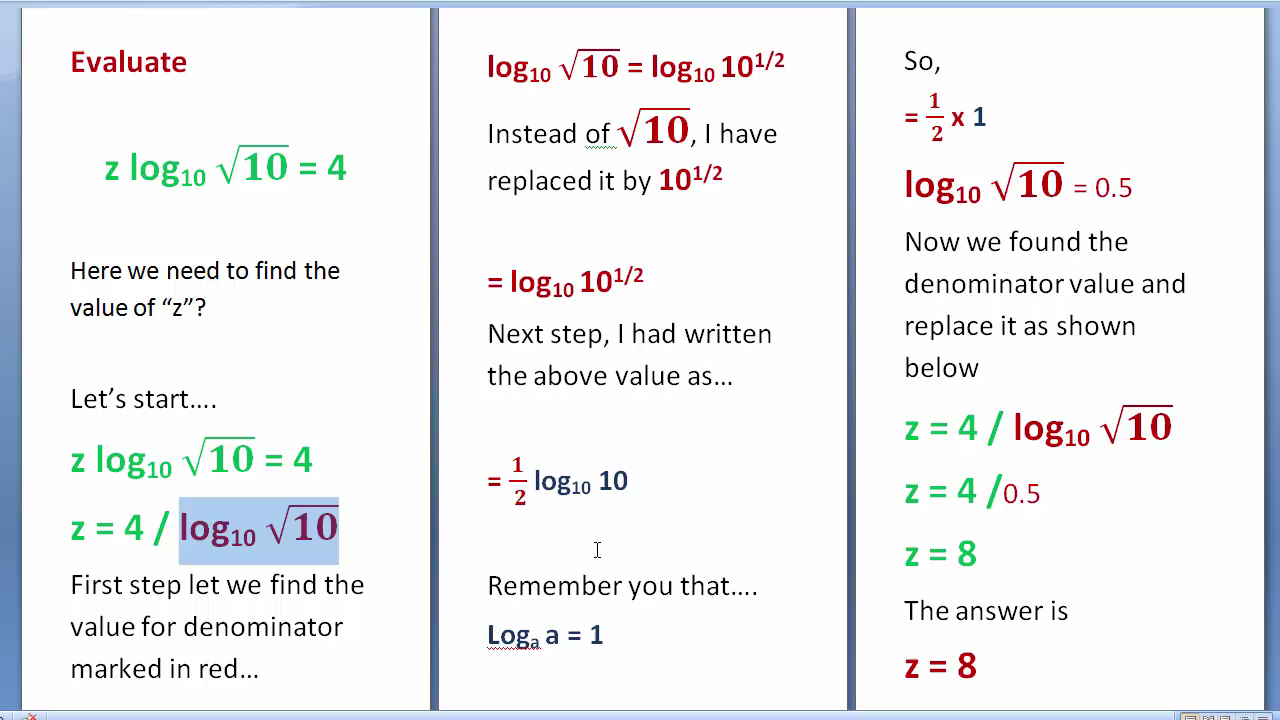}
\includegraphics[width=1.65in]{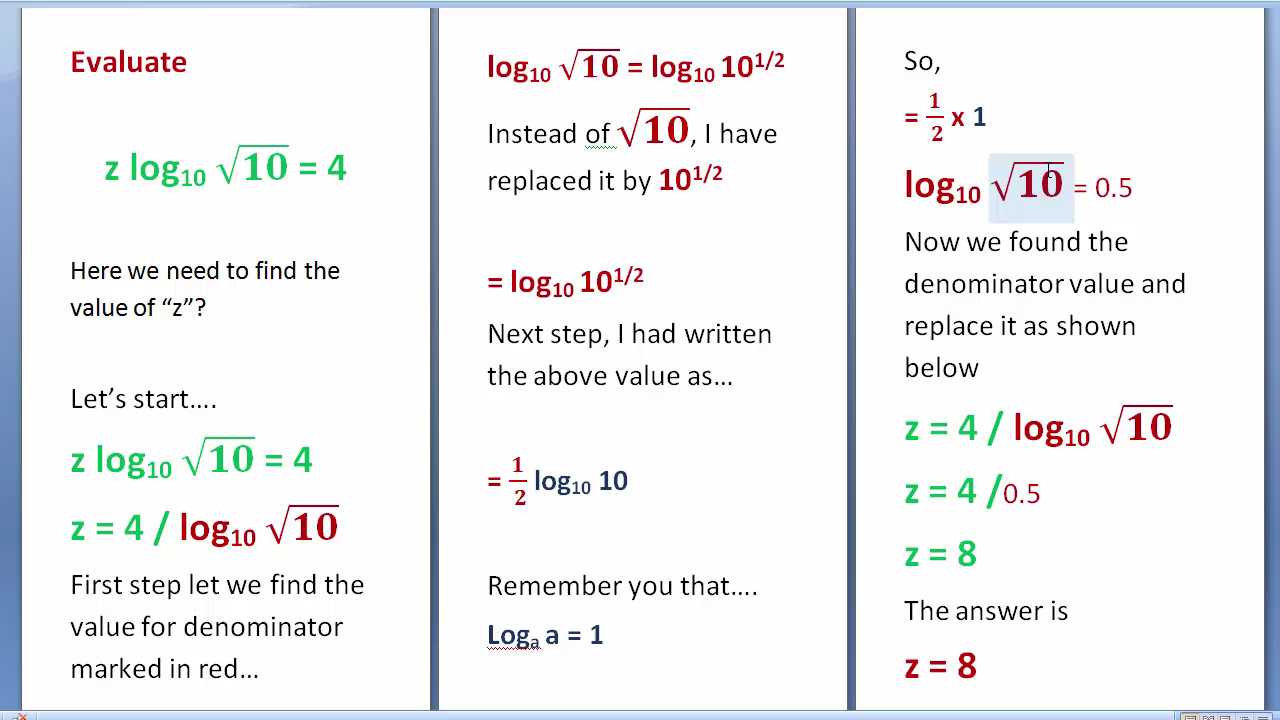}
\includegraphics[width=1.65in]{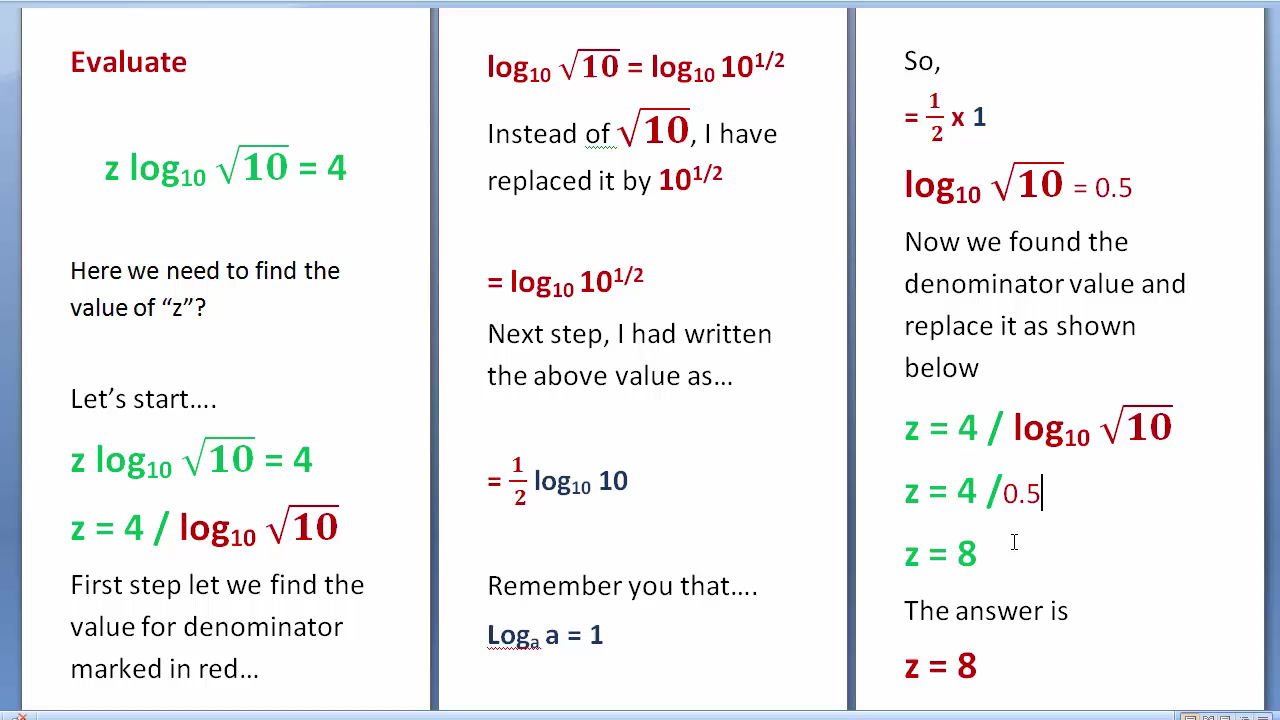}
\end{center}
    }
%
\caption{Snapshots (at 25\%, 50\%, 75\%, and 100\% time duration within each video) of 5
representative examples of 399 total crowdsourced explanatory videos on logarithms.}
\label{fig:videos}
\end{figure*}

Over 2 data collection periods consisting of approximately 2 weeks each, we collected 399 videos from 66 unique teachers
(17\% female; minimum reported age of 18, maximum reported age of 55) -- approximately 6 vidoes per participant.
This corresponds to approximately 100 videos per week of active data collection.
The duration of most videos was between 1 and 3 minutes.
Interestingly, several of the participants in our study expressed to us via email
their enjoyment in completing the HIT, and many of them created explanations for several different problems.
See Figure \ref{fig:videos} for a representative sample of the crowdsourced videos.

\subsubsection{Analysis of correctness}
We have manually annotated 145 out of these 399 videos (annotation is still ongoing) as being (a) 
proper submissions and (b) mathematically correct. In order to be considered a proper submission,
the video had to be both novel (e.g., not just copied from Youtube) and a \emph{bona fide} video (e.g., not just
a static image frame with no accompanying audio that was packaged as a video file).
In order to be considered mathematically correct, the video needed to begin with
the problem statement, end with the correct solution, and
contain no statement that was objectively false. (At this stage, we made no attempt to judge the
pedagogical effectiveness of the videos; we will explore this in Experiment II).
Finally, we labeled a video as ``borderline'' (rather than correct or incorrect) for minor missteps such as when
the teacher referred to a mathematical \emph{expression} (e.g., $\log_2 1$) as
an \emph{equation} even though there was no equals sign.

Of the 145 annotated videos, 117 ($81\%$ of 145) were judged to be mathematically fully correct;
16 videos ($11\%$) were judged as incorrect; 7 ($5\%$)  were judged as ``borderline''; and
5 ($3\%$) were not proper submissions.

\subsubsection{Examples of mistakes}
Some of the mistakes were incorrect verbal usage of terminology even if the
written derivation itself was correct. For example, one teacher read the expression $\log x$ as
``log times x'' instead of ``log of x''. Other mistakes were more egregious. For instance, in one video, the teacher
``canceled'' two occurrences of the $\log$ function -- one in the numerator and one in the denominator:
\begin{equation*}
\frac{\log \frac{1}{4}}{\log \frac{1}{2}} = \frac{\hcancel[red]{\log} \frac{1}{4}}{\hcancel[red]{\log} \frac{1}{2}} 
                                          = \frac{\frac{1}{4}}{\frac{1}{2}}
\end{equation*}
(Interestingly, his final answer to the problem -- due to another mistake -- was actually correct.)

\section{Examining the diversity of the videos}
In order for personalized learning systems to be effective for a wide variety of students, they must be able
to draw from a large and \emph{diverse} set of learning resources in order to give each individual student the kind
of help she/he needs most.
We thus performed a qualitative analysis of the crowdsourced videos for diversity
along the dimensions of presentation format, language, and pedagogical approach used to solve the math problem.

\subsection{Presentation format}
As shown in Figure \ref{fig:videos}, there was diversity in  the presentation formats and styles used
in the videos. The five most common formats include:
(1) a video of the teacher writing on paper (2) a video of the teacher's
computer screen that is used as an electronic notepad; (3) a video of the teacher speaking directly to the learner in a face video
along with written materials (sometimes held up to the camera) to show the derivation; (4) a step-by-step ``Powerpoint''-style
presentation; and (5) a static Powerpoint slide to which the instructor points using the mouse.
In addition, all videos included accompanying audio to explain the solution.
Some teachers also mixed styles by writing on the Powerpoint slide using a mouse-based paint tool.

\subsection{Language}
Although all crowdsourced videos were in English, there was variability in the geographical origin
and dialect of the spoken English. In particular, several teachers 
used terminology such as ``5 into x'' to express the
\emph{multiplication} of 5 with $x$, i.e., $5x$. This terminology is widely used in India \cite{wiki:into}. This also highlights the
need for both a large, diverse set of explanations as well as smart decision-making in determining
which learners are assigned to which explanation.

\subsection{Pedagogical style}
Over the 18 math problems for which tutorial videos were crowdsourced,
we observed two general approaches that teachers used to derive the solutions to the problems.
In some explanations, the \emph{definition of logarithm} -- i.e., the logarithm of $x$ base $b$
is the power to which $b$ must be raised to equal $x$ -- was invoked to solve the problem.
For example, to reduce $\log_{10} 1000$, one can use the fact that clearly $10^3=1000$ to arrive at
the correct answer of $3$.
In other explanations, the teacher emphasized the \emph{syntax} of logarithms and how rules can
be applied to transform a problem step-by-step into the solution.
For example, to simplify $\log_x x^4$,  the teacher would note that
$\log_x y^c = c \log_y x$ for all $c$ to derive $4 \log_x x$; then, he/she would note that
$\log_x x =1$ for all $x$ to derive $4\times 1=4$.

\subsubsection{Path Analysis}
One important way in which personalized learning systems can help students is to provide
feedback and hints that are tailored to the particular \emph{solution path} that  student took
toward finding a solution \cite{barnes2008toward}. We thus investigated whether the crowdsourced explanations
exhibited diversity in terms of the teachers' own solution paths. In particular,
for the particular math problem, ``Solve for $x$: $x \log_4 16 = 3$'', we performed a  path analysis 
in which we compared the different derivations paths that the different teachers used to arrive at a solution.
We watched each of the 17 different explanation videos that were crowdsourced from people
on Mechanical Turk and manually coded
for all equations that the teacher wrote, in the order that she/he wrote them.
Since different teachers would 
express the exact same mathematical relationships in different ways, we devised a set of 10 ``equivalency rules'' to eliminate trivial
syntactic differences. For example, one equivalency rule was 
$\log_a a=1$ is equivalent  to $\log_x x=1$. Even after applying these rules to each pedogogical path, each of the teacher's paths
was unique.

\begin{figure*}
\begin{center}
\includegraphics[width=6.5in]{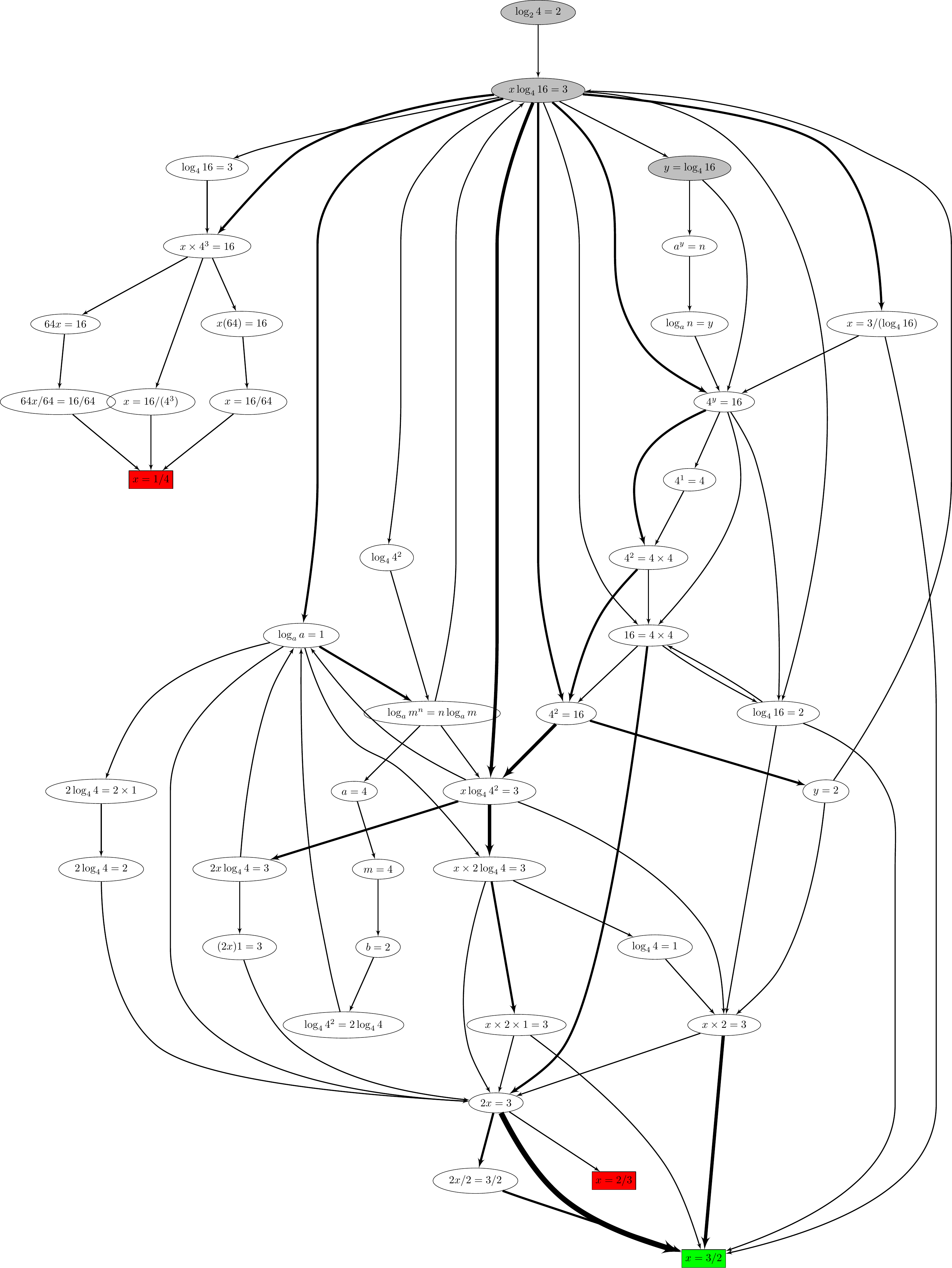}
\caption{Graph of different paths taken by 17 different explanations of how to solve the problem: ``Solve for $x$:
$x \log_4 16 = 3$.'' Graph edges are weighted by the number of teachers whose solution transitioned
from one state to another. Rectangular nodes are terminal states (i.e., the end of a solution path), where
red indicates a mathematical error and green indicates correct solution. Gray ellipsoid nodes are different start
states taken by different teachers.}
\label{fig:problem17paths}
\end{center}
\end{figure*}

To represent visually the collection of all paths, we constructed a graph (see Figure \ref{fig:problem17paths})
whose nodes consisted of the union over all teachers of the states reached by their pedagogical paths. The weight of each directed
edge in the graph corresponded to
the number of teachers whose path transitioned from one state to another.
Rectangular graph nodes are terminal states (i.e., the end of a solution path), where
red indicates an incorrect solution (e.g., $x=1/4$) and green indicates correct solution ($x=3/2$).  Gray ellipsoid nodes are the start
states taken by different teachers. While most teachers started with the problem statement, a few teachers did not; e.g.,
one teacher first introduced an easier logarithm calculation ``$\log_2 4 = 2$'' before returning to the actual problem.

{\bf Results}: Although there was substantial overlap in the teachers' solution paths,
all 17 of them were unique -- see Figure \ref{fig:problem17paths}.
In particular, we observe several dimensions of variability in teachers' pedagogical approaches:
\begin{enumerate}
\item {\bf Strategy}: Some teachers applied syntactic laws of logarithms to
derive their solution; such solutions passed through nodes in the lower left quadrant of the graph (e.g.,
$\log_a a=1$, $\log_a m^n = n \log_a m$). Other teachers 
appealed to the definition of logarithm to infer, for example, that $\log_4 16 =2$
(since $4^2=16$ -- see upper right quadrant of graph).
\item {\bf Sequence}: Paths that passed through the same states varied in the order in which states were visited.
For example, some paths presented $\log_a a=1$ before $\log_a m^n =n \log_a m$, and some paths did the reverse.
\item {\bf Granularity}: Some paths contained considerably more detail than others. For example, in paths
that tackled the sub-problem of determining what value $y$ solves $4^y=16$, some paths first provided
simple arithmetic examples of exponentiation ($4^1=4$, $4^2=4\times 4$, etc.), whereas other jumped directly
to the answer $\log_4 16=2$).
\end{enumerate}

\section{Experiment II: Finding the Best Videos}
\label{sec:learning_gains}
Experiment I revealed that the majority ($81\%$) of the submitted videos were both proper submissions (i.e., were novel \emph{bona fide} videos)
and were mathematically correct. In this section, we explore whether the videos show any promise for actually helping students to learn.
Because this study is about crowdsourcing novel explanations from ordinary people
around the world who may have varying mathematical skill and pedagogical expertise, we do not expect \emph{all}
the videos to be effective in helping students to learn. Rather,  we assessed whether 
the \emph{average} learning effectiveness of the videos -- quantified  by posttest-minus-pretest score of participants who watched the
videos in a separate experiment -- was statistically significantly
higher than the learning effectiveness of a ``control'' video about a math topic unrelated to logarithms 
(specifically, a historical tutorial about the number $\pi$).

With this goal in mind, we randomly sampled 40 videos
from the 117 that were confirmed (out of the 145 total that were annotated) to be mathematically correct
and conducted an experiment in which each participant took a pretest on logarithms, watched
a randomly assigned video (either one of the 40 crowdsourced videos or the control video),
and then took a posttest. In contrast to Experiment I, the participants in this experiment
were not expected to know anything \emph{a priori} about logarithms.

\subsection{Participants}
We recruited $N=200$ participants from Amazon Mechanical Turk.
Each participant who completed the experiment received $\$0.40$ payment.

\subsection{Apparatus}
We created a Web-based pretest on logarithms using the problems shown in Figure \ref{fig:problems},
and also a posttest whose content was similar in length, content,
and difficulty to the pretest but contained different problems. The pretest and posttest
were borrowed from the study in \cite{salamanca2012characterizing}.

\subsection{Procedure}
The experiment proceeded as follows:
\begin{enumerate}
\item The participant took the pretest.
\item The participant was randomly assigned a video to watch. With probability $0.2$, the participant
was assigned the control video, and with uniform probability of $0.8/40=0.02$, the participant was assigned to watch one of the
40 crowdsourced videos.
\item The participant took the posttest.
\end{enumerate}

\subsection{Dependent variables}
The dependent variables  in this experiment were the average learning gains
\[ G_k \doteq \frac{1}{|V(k)|} \sum_{i\in V(k)} (\textrm{post}_i - \textrm{pre}_i) \]
for each video $k$,
where $\textrm{pre}_i$ and $\textrm{post}_i$ are the pretest and posttest scores for participant $i$, and $V(k)$ is the set of participants
who were assigned to watch video $k$.

\subsection{Results}
\begin{figure}
\begin{center}
\includegraphics[width=3.35in]{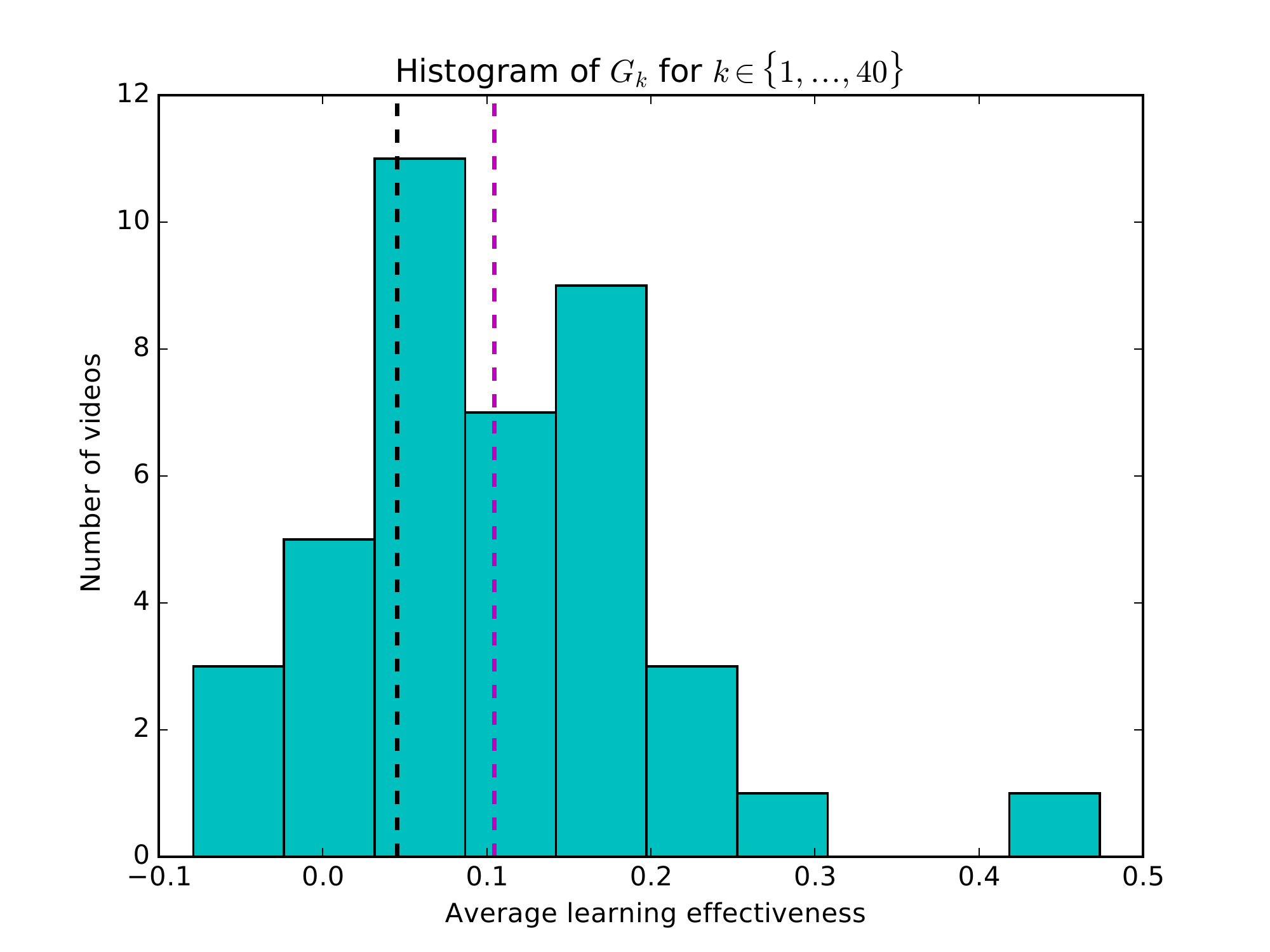}
\caption{Histogram of the average learning gains $G_k$ (average posttest minus pretest score across all
subjects who watched video $k$)
for the 40 ($k\in \{1,\ldots,40\}$) crowdsourced videos.
The black dashed line shows the average learning gains for the ``control'' video; the magenta line shows
the average learning gains of all the crowdsourced videos.}
\label{fig:learning_gains_dist}
\end{center}
\end{figure}

The histogram of the $G_k$ for $k\in \{1,\ldots,40\}$ is shown
in Figure \ref{fig:learning_gains_dist}. The average learning gains ($0.105$) for the
$40$ crowdsourced videos was higher than for the control video ($0.045$); the difference was statistically
significant ($t(39) = 3.715$, $p<0.001$, two-tailed).

\subsection{Differential Drop-out}
Since some subjects started but did not complete the experiment, the number of subjects
collected per video varied. This issue of differential drop-out can lead to distorted estimates:
for example, if one tutorial video is particularly bad and only those students who are already proficient in logarithms
decide to persist through the bad video and complete the HIT, then the estimated learning gains for that video
might be positively biased. Unfortunately, Amazon Mechanical Turk does not provide an easy mechanism to track
which workers started, but did not complete, the experiment -- data are available only for participants
who finished the post-test and submitted the HIT. However, since we do know how many participants completed
the HIT for each video, and since we know the prior probability of assigning each participant to each video,
we can assess whether some videos resulted in drop out more often than others.
Specifically, we conducted a Pearson's
$\chi^2$ test where the vector of probabilities for the 41 videos (1 control plus 40 crowdsourced videos)
was $[ 0.2\ \frac{0.8}{40}\ \ldots\ \frac{0.8}{40} ]$. 
The result of the test ($\chi^2(40)=34$, $p=0.7363$)
indicate that the  \emph{completion} rates for the videos 
were not statistically significantly different from the \emph{assignment} rates.
Though this result does not mean that the  estimates of learning effectiveness
in Figure \ref{fig:learning_gains_dist} are completely unbiased, it provides
some evidence that they are not to be completely discounted.


\section{Experiment III: Comparing to Khan Academy}
In our third experiment, we compared the learning gains of the best 4 videos as estimated in Experiment II,
to the learning gains of a popular tutorial video on logarithms produced by Khan Academy
(specifically \url{https://www.youtube.com/embed/Z5myJ8dg_rM}, with $924,520$ views as of October 20, 2016).

\subsection{Participants}
We recruited $N=250$ participants from Amazon Mechanical Turk.
Each participant who completed the experiment received $\$0.40$ payment.

\subsection{Apparatus}
Same as in Experiment II.

\subsection{Procedures}
Same as in Experiment II, except that each participant was assigned uniformly at random to watch
one of five different tutorial videos: 4 of these videos were crowdsourced videos, and 1 was the Khan Academy video.

\subsection{Dependent variables}
Same as in Experiment II.

\subsection{Results}
\begin{table}
\begin{center}
\begin{tabular}{c|c|l}
Video  & Participants & $G_k$ \\\hline
1 & 58 & 0.1416 \\
2 & 42 & 0.1140 \\
3 & 57 & 0.0942 \\
4 & 35 & 0.0932 \\
Khan & 58 & 0.1506 \\
\end{tabular}
\end{center}
\caption{Average learning gains $G_k$ as measured in Experiment III, for the 4 videos
were estimated to be highest in Experiment II, compared to the average learning gains of
a popular Khan Academy video on logarithms.}
\label{tbl:results_experiment3}
\end{table}
As shown in Table \ref{tbl:results_experiment3}, the learning gains associated with the Khan Academy video compared to the best of the
4 crowdsourced videos were very similar -- $0.1506$ versus $0.1416$, respectively. The difference between them was not
statistically significant ($t(114)=0.2277,p=0.82$, two-tailed).

We note the following issues when comparing the crowdsourced math videos to the Khan Academy video: On the one
hand, the Khan Academy video was substantially longer
(7 minutes and 2 seconds) than the  4 crowdsourced videos  (maximum length 2 minutes and 16 seconds) and hence can contain subtantially
more potentially useful math content. On the other hand, the content presented in the crowdsourced videos 
was arguably more closely aligned to the post-test (though none of the questions explained in the video was exactly the same
as any problem
on the post-test) than was the Khan Academy video. Nonetheless, the results suggest that math tutorials crowdsourced
from ordinary people on the Web can, at least sometimes, produce high-quality educational content.


\section{Results \& Conclusions}
In the study described in this paper we explored how to devise a crowdsourcing task for use on Amazon Mechanical
Turk in which ordinary people are asked to take on the role of a ``teacher'' and create novel tutorial videos
that explain how to solve specific math problems related to logarithms. Further, we analyzed qualitatively the crowdsourced
videos for mathematical correctness, diversity across several dimensions including pedagogical approach, presentation
format, and language style. Finally, we assessed the utility of the best such videos in terms of helping students
to learn, measured as posttest minus pretest performance by students who were asked to watch one of the
crowdsourced videos in a separate experiment.

Results from this study suggest that: (1) Crowdsourcing of
full-fledged tutorial videos from ordinary people
is feasible, provided that appropriate guidelines (e.g., about using clear handwriting) on how to craft the explanations are provided.
In fact, several of the crowdsourced workers expressed enthusiasm for the task, which likely requires more creativity
than the kinds of tasks that are typically crowdsourced (e.g., image tagging). (2) Crowdsourcing from a 
large number of ``teachers'' (66 teachers collectively produced 399 videos in our study) can produce a set
of learning resources that exhibits considerable diversity along the dimensions listed above. (3)
Although a few of the crowdsourced tutorial videos -- which would need to be filtered out -- contained important mathematical errors,
the best of these videos were statistically significantly more effective, in terms of helping students to learn,
than what would be expected from a ``control'' video on an irrelevant math topic. In fact, in terms of associated
learning gains, the very best crowdsourced
videos were comparable -- and statistically indistinguishable from -- a popular tutorial video on logarithms produced
produced by Khan Academy. In sum, these findings provide support for the hypothesis that crowdsourcing
can play an important role in collecting large, rich, and diverse sets of educational resources that enable personalized learning at scale.

{\bf Future work} within this project will investigate machine learning-based
methods (e.g., \cite{williams2016axis,mooclets2015,lancontextual}) to inferring which students should receive which tutorial videos --
based on joint properties of students and teachers -- in order to maximize their learning gains. Moreover, computer vision
techniques based on deep neural networks will be explored in order to facilitate large-scale, automatic
annotation of learning resources for important characteristics -- such as pedagogical approach and presentation style --
that can be used to recommend specific resources  to specific learners.


\section*{Appendix}
\begin{figure*}
\noindent
    \parbox{\textwidth}{
    \noindent\fbox{%
    \parbox{\textwidth}{
\large \textbf{Consent Form \& Video Recording Release Form}\\
\small ... You will then be asked to create a novel video in which you explain how to solve a short mathematical exercise: PROBLEM. The content and format of the video are up to you, but the video must address the problem and must be mathematically correct. For example, the video might contain a screencast showing an electronic ``blackboard'' on which you explain how to answer the problem. Alternatively, you might prefer to talk into a web camera and record a video of your face and your voice. ...
    }}

    \noindent\fbox{%
    \parbox{\textwidth}{
\large \textbf{Survey}\\
\small
	Please answer the questions below. When you are done, click "Next".

	\begin{enumerate}
	\setlength\itemsep{0em}
	\item How old are you (in years)? 
	\item What is your gender? 
	\item What is the highest level of education you have completed?
		\ldots
	\item How much do you enjoy mathematics?
		\ldots
	\end{enumerate}
    }}

    \noindent\fbox{%
    \parbox{\textwidth}{
\large \textbf{Sample Problems \& Explanations}\\
\small
	This page contains some example videos that explain how to solve math problems.
        Please watch the videos carefully so you know what we are looking for in this HIT.\\
	\includegraphics[width=.8in]{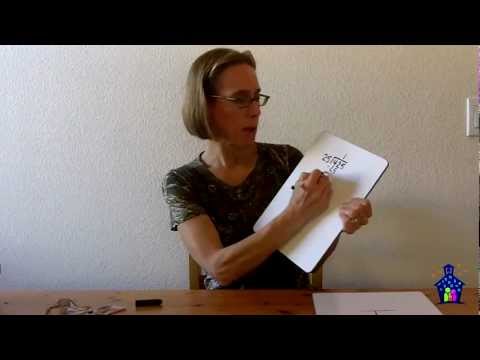}
	\includegraphics[width=.8in]{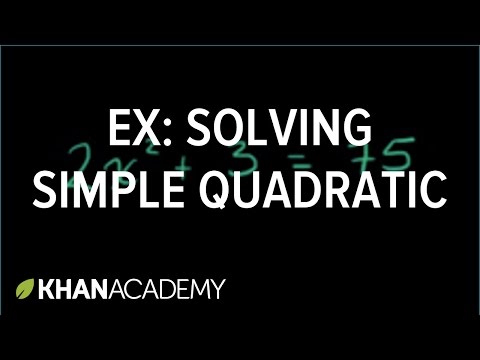}
	\includegraphics[width=.8in]{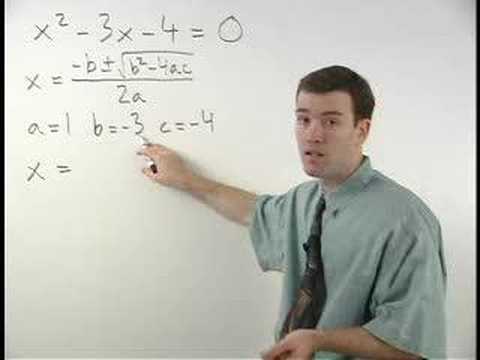}
    }}

    \noindent\fbox{%
    \parbox{\textwidth}{
\large \textbf{Hints on Making a Good Video}\\
\small
	When you make your video, you may sometimes record images of your own handwriting.
	Please look at the following handwriting examples so you know what distinguishes a good video
	from a bad video. Note that a bad video may be rejected due to poor image quality. 
	
	The following 2 examples are \textbf{OK} -- the writing is dark, big, and clear.\\
	\includegraphics[width=1.75in]{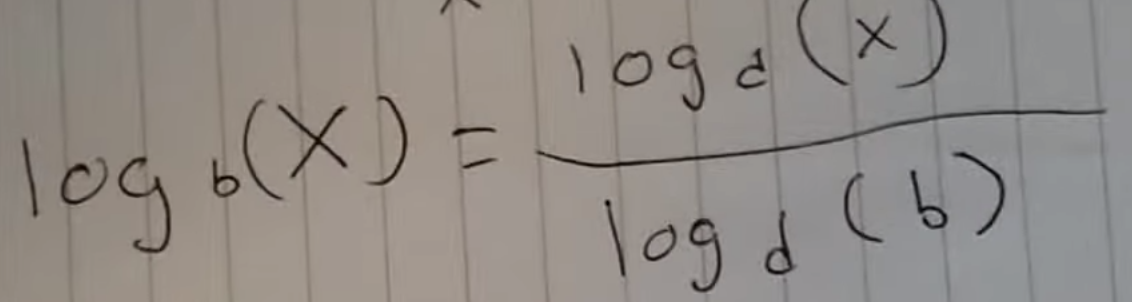}
	\includegraphics[width=1.75in]{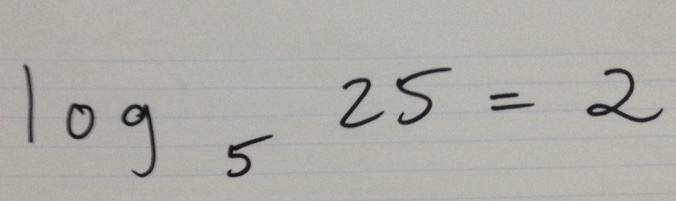}

	The following 3 examples are \textbf{not OK} -- the writing is too small, blurry, and/or hard to read.\\
	\includegraphics[height=.8in]{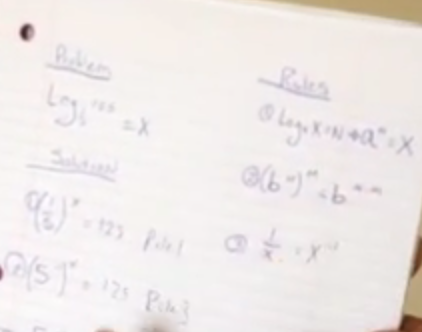}
	\includegraphics[height=.8in]{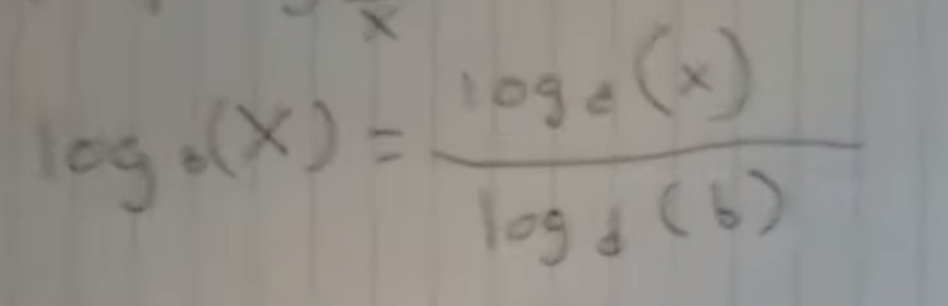}
	\includegraphics[height=.8in]{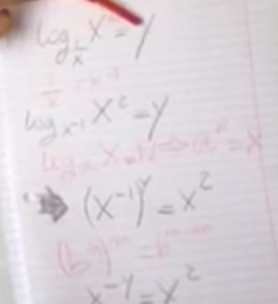}
    }}

    \noindent\fbox{%
    \parbox{\textwidth}{
\large \textbf{Problem \& Instructions}\\
\small
	Please examine the following math problem: PROBLEM

	\noindent {\bf Instructions}:
		\begin{enumerate}
		\setlength\itemsep{0em}
		\item Think carefully about how you would explain to someone else how to solve this problem.
		\item Create a video that explains how to solve the problem.
		\item Upload the video to our server.
		\end{enumerate}
	\noindent {\bf Rules}:
		\begin{itemize}
		\setlength\itemsep{0em}
		\item Your video must explain how to answer the following math problem: PROBLEM
		\item Your video must be original - it cannot be an existing video.
		\item Your video must be mathematically correct.
		\item Your video may not contain any images of children (less than 18 years old).
		\item Your video may not contain any nudity or profanity.
		\end{itemize}
    }}

    \noindent\fbox{%
    \parbox{\textwidth}{
\large \textbf{Submission}\\
\small
	...
    }}
}
\normalsize
\caption{The different screens of the Human Intelligence Task (HIT) posted to Amazon Mechanical Turk to crowdsource explanations from amateur
``teachers'' in Experiment I.
}
\label{fig:HIT}
\end{figure*}

Figure \ref{fig:HIT} shows a synopsis of the most important content of our HIT (which was rendered in HTML).

\bibliographystyle{abbrv}
\bibliography{paper}  

\end{document}